\documentclass[english,aps,prl,nofootinbib,twocolumn,superscriptaddress]{revtex4-2}

\usepackage{amsmath,amsthm,latexsym,amssymb,amsfonts,epsfig}

\usepackage{float}
\usepackage{upgreek}

\theoremstyle{remark}

\theoremstyle{remark}

\theoremstyle{obser}

\theoremstyle{obser}

\usepackage{xcolor}
\usepackage{amsmath}
\usepackage{amsthm}
\usepackage{graphicx}
\usepackage{amssymb}
\usepackage{dsfont}
\usepackage{color}
\usepackage{soul}
\usepackage[hyperfootnotes=true,unicode=true, 
 bookmarks=true,bookmarksnumbered=false,bookmarksopen=false,
 breaklinks=false,pdfborder={0 0 1},backref=false,colorlinks=true]
 {hyperref}
 \usepackage{float}
 \usepackage[markup=nocolor]{changes}

\usepackage[makeroom]{cancel}

\usepackage{etoolbox}

\usepackage{hyperref}
\hypersetup{
    colorlinks,%
    citecolor=blue,%
    filecolor=blue,%
    linkcolor=blue,%
    urlcolor=blue
}

\newcommand{\be}{\begin{equation}}
\newcommand{\ee}{\end{equation}}
\newcommand{\ba}{\begin{eqnarray}}
\newcommand{\ea}{\end{eqnarray}}
\def\bal#1\eal{\begin{align}#1\end{align}}

\makeatletter
\makeatletter
\@ifundefined{textcolor}{}
{%
 \definecolor{BLACK}{gray}{0}
 \definecolor{WHITE}{gray}{1}
 \definecolor{RED}{rgb}{1,0,0}
 \definecolor{GREEN}{RGB}{0,204,0}
 \definecolor{BLUE}{rgb}{0,0,1}
 \definecolor{CYAN}{cmyk}{1,0,0,0}
 \definecolor{MAGENTA}{cmyk}{0,1,0,0}
 \definecolor{YELLOW}{cmyk}{0,0,1,0}
 }

\usepackage{hyperref}
\hypersetup{
    colorlinks,%
    citecolor=blue,%
    filecolor=blue,%
    linkcolor=blue,%
    urlcolor=blue
}

\makeatother

\begin{document}

\title{Quantum speed limit and stability 
of coherent states in quantum gravity}

\author{Klaus Liegener}
\email{klaus.liegener@wmi.badw.de}
%\affiliation{Institute for Quantum Gravity, Friedrich-Alexander University Erlangen-N\"urnberg, Staudtstra\ss e 7, 91058 Erlangen, Germany}
\affiliation {Walther-Mei\ss ner-Institut, Bayerische Akademie der Wissenschaften, 85748 Garching, Germany}

\author{\L ukasz Rudnicki} 
\email{lukasz.rudnicki@ug.edu.pl}
\affiliation{International Centre for Theory of Quantum Technologies (ICTQT), University of Gda{\'n}sk, 80-308 Gda{\'n}sk, Poland}
\affiliation{Center for Theoretical Physics, Polish Academy of Sciences, Al. Lotnik{\'o}w 32/46, 02-668 Warsaw, Poland}

\date{\today{}}

\begin{abstract}
Utilizing the program of expectation values in coherent states and its recently developed algorithmic tools, this letter investigates the dynamical properties of cosmological coherent states for Loop Quantum Gravity.
To this end, the Quantum Speed Limit is adapted to Quantum Gravity, yielding necessary consistency checks for any proposal of stable families of states. To showcase the strength of the developed tools, they are applied to a prominent model: the Euclidean part of the quantum scalar constraint. We report the variance of this constraint evaluated on a family of coherent states showing that, for short times, this family passes the Quantum Speed Limit test, allowing the transition from one coherent state to another one. 
\end{abstract}

\maketitle
%
%While observations indicate our current understanding of both Quantum Field Theory and General Relativity to be incomplete, a theory of Quantum Gravity (QG) resolving those issues is not yet finished. It is mandatory to push suitable candidates of QG towards extracting their semi-classical limit and match it with observable data. A promising approach comes in the form of Loop Quantum Gravity (LQG) \cite{Thi:07} with its community striving actively towards deriving its cosmological sector.\\
Spectacular results concerning \textit{reduced} quantization of isotropic flat cosmology, so called Loop Quantum Cosmology (LQC) \cite{Bojo:05}, have been obtained in last decades: For sharply peaked coherent states, %defined appropriately,
the initial singularity of isotropic Universe is resolved \cite{Bojo:01} and replaced by a Big Bounce \cite{Ash:06,APS:06_b,APS:06_c,ADLP:18}. While the peak of such a coherent state follows a trajectory in classical phase space  \cite{Ding:08}, quantum effects, e.g. relevant for the power spectrum of the cosmic microwave background \cite{Agullo:12}, are also there.

However, we are left with an important open question, namely, whether a successful quantization of \textit{full} General Relativity would support the aforementioned predictions relevant for LQC. While such a complete theory is still not available, an increasing computational power allows to probe complicated lattice quantum systems \cite{CBHLQDH:20}.  A suitable construction of coherent states on the lattice already exists \cite{Thi:00}. Its variant, called gauge (cosmological) coherent states (GCS), seems useful as a litmus paper for LQC. While the Big Bounce is reproduced, deviations from LQC-modified Friedmann equations occur \cite{DL:17a}, enabling comparison with discretized General Relativity \cite{Kaminski:20}. 

Despite recent progress \cite{DL:17b,HSZ:20}, the dynamics of GCS is poorly understood. In this letter we partially address that important aspect by virtue of {\it quantum speed limit}  \cite{Mandelstam1991,Fleming,Aharonov,Zych} --- a tool familiar from fundamental considerations concerning time-energy uncertainty relations in quantum mechanics. We reformulate this tool to study ``stability" of an arbitrary single-parameter family of "target" states $\Phi_s$, given a generator of evolution $\hat H$. We call such a family stable with respect to  $\hat H$, if  $\Phi_s\equiv e^{-i s \hat H}\Phi_0$. Afterwards, we touch upon the implementation of the constraints in the quantum domain to argue why stability is important in Loop Quantum Gravity (LQG). Finally, we employ the quantum speed limit, unravelling properties of the effective behaviour of the family of cosmological coherent states.

\textit{Quantum speed limit as consistency check for stability.}---In our setting, the state $\Phi_0$ is both an initial state of evolution rendered by the generator $\hat H$, as well as a member of a given family of states $\Phi_s$. However, suspecting that the evolved state $\Phi(s):=e^{-i s \hat H}\Phi_0$ does not belong to the family in question, i.e. $\Phi(s)\neq\Phi_s$, we wish to quantify its deviation from $\Phi_s$. To this end we introduce a two-point quantum fidelity
\begin{align}\label{eq:Wapprox1}
    W(s,\tau) := |\langle \Phi(s) , \Phi_{\tau} \rangle|^2.
\end{align}
Note that we label time-evolution by $s$ and $\tau$, not following the standard notation with $t$. We do this for the sake of further consistency with the literature on complexifier coherent states in LQG, where $t$ indicates the spread of a state.

For our purpose, we present a slightly more general formulation of quantum speed limit (QSL):
\begin{align}\label{qsl}
\tau\cdot\Delta_{\Phi_0} \hat H /\hbar \geq  | A(0,\tau) - A(\tau,\tau) |,
\end{align}
\begin{align}
    A(s,\tau):=\arccos\left(\sqrt{W(s,\tau)}\right),
\end{align}
derived in the first section of Supplementary Material. As usual, $\Delta_\Psi \hat{O}$ is a standard deviation of an observable $\hat O$, given a state $\Psi$. In a typically considered scenario in which stability is given, one just sets  $A(\tau,\tau)=0$. In addition, assuming that the initial and the final states are orthogonal, i.e. $W(0,\tau)=0$, the bound in (\ref{qsl}) equals $\pi/2$. In this way, QSL helps understand how quickly the transition between two (orthogonal) states can happen due to $\hat{H}$. 

On the contrary, we aim to test whether it is even legitimate to hope for stability, i.e.  $W(\tau,\tau)\approx 1$. To this end, using a basic inequality $|x-y|\geq |x|-|y|$, we obtain an equivalent formulation of (\ref{qsl})
\begin{align}\label{qslrev}
|A(\tau,\tau)|  \geq | A(0,\tau)| - \tau\cdot\Delta_{\Phi_0}\hat{H}/\hbar.
\end{align}
Therefore, a necessary condition for stability is that the right hand side does not become \textit{too} positive.
Before we apply the above consistency check to Quantum Gravity directly, we shall first ask why stability of coherent states is important in quantization of fully constrained theories (i.e. where the Hamiltonian is a linear combination of constraints) as happens in the case of General Relativity.

\textit{Classical, strong and weak implementation of constraints.}---Let $C_i$ denote the constraints of the theory\footnote{For General Relativity, $C_i$ constitutes out of scalar-, diffeomorphism- and Gauss constraint as $C_i$.}. In the classical scenario, one can speak about a valid solution only if all $C_i$ are satisfied (i.e. vanish) at a given point of the phase space. How this rule is supposed to be carried over to the quantum level remains an open debate. We recall two known strategies and propose a third option.

First of all, one can avoid the problem by implementing the constraints already at the classical level. Most prominently, in General Relativity one has introduced so called deparametrized models \cite{VP:12}, where diffeomorphism and scalar constraints are fixed on the classical level via dust fields. Afterwards, dynamical evolution is described by a true Hamiltonian\footnote{Albeit not custom, it is possible to similarly implement the Gauss constraint classically, at the expense of a more complicated phase-space structure.}. In this case the question for stable coherent state becomes immediately interesting and asks for applying the consistency check of the QSL.

Alternatively, implementation of some $C_i$ on the quantum level can be sought. Dirac's quantization proposal suggests to find the kernel of the constraint operator $\hat{C}_i$ and denote it as the {\it physical} Hilbert space. Such a \textit{strong} implementation unfortunately suffers from severe obstacles due to the scalar constraint, since its kernel has not been found so far. Therefore, we consider \textit{weak} implementation which admits two ways of relaxation, in comparison with the strong implementation. The constraints must hold (at least) on average, and only up to a certain quantum tolerance scale ``$t$". More formally and given some $C_i$, the state $\Psi$ is called {\it physical} if it is sharply peaked:
\begin{align}
\langle \hat{C}_i  \rangle_\Psi = 0 +\mathcal{O}(t), \quad \Delta_\Psi\hat{C}_i =0 +\mathcal{O}(\sqrt{t}),
\end{align}
Weak implementation of constraints admits that associated symmetries are ultimately broken in the quantum theory, however at the macroscopic level this violation would be extremely difficult to detect. We note in passing that a similar weak implementation has already been suggested for the so-called simplicity constraint of higher dimensional LQG \cite{GLY:19}.  

Physical states subject to weak implementation of the constraints are not automatically invariant, i.e. in general $\Psi\neq\exp(i s \hat{C}_i)\Psi$. This leaves room for two possibilities. In the first one,  only gauge-invariant observables $\hat O$ are considered while describing physically-relevant quantities. Such observables must commute with the quantum constraints, so that $\hat O =e^{is \hat C_i} \hat O e^{-is \hat C_i}$ holds. A prime candidate for this is the quantum Gauss constraint of Lattice Gauge Theories, as constructing observables commuting with it is well understood\footnote{Hence, when using the GCS (\ref{eq:coh_state_def}) later on, no further corrections appear from the quantum Gauss constraint. E.g., Dirac's quantization would require to change the coherent states which are not in the kernel of the quantum Gauss constraint. Implementing it strongly amounts to an additional term in the next-to-leading order corrections of any expectation value. For further details see the appendix of \cite{DL:17b}.}. An alternative to gauge-invariant observables (especially when their construction is difficult, e.g. for the scalar constraint) can be the aforementioned weak implementation, in  which one would ask for coherent states that are {\it stable} in the sense of following a sharply peaked trajectory mapping in classical phase space the associated gauge orbit.

It transpires that stable coherent states are of paramount importance for the approach of weak implementation for complicated constraints as well as for the deparametrization approach, where the constraints get replaced by a usual evolution operator. While proposals for coherent states exist in Quantum Gravity, they have so far never been investigated for their stability. This sets the motivation to apply the previous consistency check via the quantum speed limit to LQG as we will do now.

\textit{Quantum gravity on a lattice.}---We recall the basic setup of Loop Quantum Gravity (LQG) on a cubic lattice, the latter composing the discretized, compact spatial slices $\sigma$ of a  4-dim manifold. Passing from classical field theory to quantum theory with finitely many degrees of freedom (with suitable infra-red regulator) is similar to considering Lattice Gauge Theories, albeit, with a few crucial differences: LQG rests on the insights of Sen, Ashtekar, Barbero et al. \cite{Thi:07,Ash:86}  that gravity can be understood as ${\rm SU}(2)$-gauge theory with canonical pair $E^I_a(x)$, $A^a_I(x)$ for $x\in\sigma$, subject to the standard diffeomorphism and scalar constraints and an additional Gauss constraint.
%\begin{align}\label{eq:DiffeoCon}
%    D_a(x)=\frac{2}{\kappa \beta}(F^J_{ab} E^b_J-G_J A^J_a)
%\end{align}
%and the {\it scalar constraint}
%\begin{align}\label{eq:ScalarCon}
%    &C= C_E(x)-(1+\beta^2)\frac{s}{\kappa}K^{[M}_aK^{N]}_b\frac{E^a_M E^b_N}{\sqrt{|\det(E)|}}\\
%      &C_E=\frac{s}{\kappa}F^J_{b}\epsilon_{JKL}\frac{E^a_K E^b_L}{\sqrt{|\det(E)|}}
%    \,.\label{eq:EuclPart}
%\end{align}
%with $s:={\rm sgn}(\det(E))$,  $\kappa$ being the gravitational coupling constant, $F^I$ the curvature of the connection and $K^I$ the extrinsic curvature.
%The kinematical phase space of ${\rm SU}(2)$ Yang-Mills theory must further be subject to the  {\it Gauss constraint}
%\begin{align}\label{eq:GaussCon}
%    G_I(x)=\partial_a E^a_I+\epsilon_{IJK} A^J_a E^a_K
%\end{align}
%to become equivalent to General Relativity.\\
Canonical quantization promotes the kinematical phase space to the Hilbert space $\mathcal{H}$, which has tensor-product structure over edges $e$ of a cubic lattice $\gamma$ with $M$ many vertices in each direction (employing periodic boundary conditions):
\begin{align}
    \mathcal{H}=\bigotimes_e \mathcal{H}_e,\hspace{30pt} \mathcal{H}_e= L_2 ({\rm SU}(2), d\mu_H),
\end{align}
where, for each edge, $\mathcal{H}_e$ is the space of square-integrable (with Haar measure $\mu_H$) ${\rm SU}(2)$-valued functions.
%Due to the Peter-Weyl theorem a possible basis on each $\mathcal{H}_e$ consists of the Wigner-D matrices:
%\begin{align}
%    \psi(g)= \sum_{jmn} c_{jmn} D^{(j)}_{mn}(g) \Leftrightarrow \psi \in \mathcal{H}_e
%\end{align}
%with $g\in{\rm SU}(2)$ and $c_{jmn}\in\mathbb R$.
Let $f_e\in \mathcal{H}_e$, then the basic operators defined on $\mathcal{H}_e$ are the {\it holonomy} operators

\begin{align}\label{eq:holonomy}
    (\hat{h}^{(j)}_{mn}(e) f_e)(g) =  D^{(j)}_{mn}(g)f_e(g),
\end{align}
with $D^{(j)}_{mn}(g)$ being a Wigner-matrix of a group element $g$ in the $2j+1$-dim, irreducible representation of ${\rm SU}(2)$, and the {\it gauge-covariant flux} operators
\begin{align}\label{eq:flux}
    (\hat{P}^K(e) f_e)(g)=-\frac{i \hbar\kappa\beta}{2} (R^K f_e)(g),
\end{align}
with, following the standard notation of \cite{Thi:07}, $\kappa=8\pi G$ being the gravitational coupling constant, $\beta>0$ being the Immirzi parameter, $\hbar$ being Planck constant and $c=1$. Finally, the {\it right-invariant vector fields} are $R^K $ and their index $K$ labels a basis of $\mathfrak{su}(2)$.
With the above structure it is possible to promote a discretization of the constraints to quantum operators, albeit current proposals are plagued by discretization artefacts of the order of the finite lattice spacing $1/M$.

\textit{Cosmological coherent states.}--- To obtain explicit expressions for the variances in order to apply the QSL to LQG, we rely on the extraction of next-to-leading-order quantum corrections with respect to the ``semi-classicality" parameter $t\geq 0$, previously referred to as the quantum tolerance scale. As an input we take the GCS \cite{Thi:00} sharply peaked over flat isotropic FLRW Universe on a torus (the classical field configuration $G\in {\rm SL}(2,\mathbb{C})$), with their building blocks given by
\begin{align}\label{eq:coh_state_def}
    \psi^t_{G} (g) := \sum_{j\in \mathbb{N}_o/2}(2j+1) e^{-t ((2j+1)^2-1)/8} D^{(j)}_{mm}(G \, g^\dagger).
\end{align}
Here, we chose a lattice oriented along the coordinate axes of the torus and for each edge going in the same direction we pick the same element $G_I=n_I e^{- z\tau_3} n_I^\dagger$, where $z=\xi -i \eta$ with $\xi,\eta\geq 0$, while $n_I$ is such that $\tau_I = n_I \tau_3 n_I^\dagger$ where $\tau_I=-i \sigma_I/2$ with $\sigma_I$ being the Pauli matrices. Finally, we define the {\it bare} isotropic, flat cosmology coherent state as the tensor product:
\begin{align}\label{eq:CCS}
    \Psi^t_{z}(\{g\}):= \langle1\rangle^{-3M^3/2}\prod_{I\in \{1,2,3\}}\prod_{k\in\mathbb{Z}^3_M} \psi^t_{G_I}(g_{k,I}),
\end{align}
where the norm $\langle 1\rangle$ is the same on every edge \cite{DL:17b} and $\mathbb{Z}_M=\{1,2,...,M\}$. These normalized states enjoy many desired properties of GCS (see \cite{Giesel:06,DL:17a,DL:17b,LZ:19}), in particular, they are sharply peaked for any observable $\hat{O}$ being a polynomial in the basic operators, in the sense that
\begin{align}\label{eq:spread}
    \left(\Delta_{\Psi^t_{z}} \hat{O}\right) / \langle \hat{O}\rangle_{\Psi^t_{z}} =\mathcal{O}(t).
\end{align}
Moreover, %their overlap is explicitly known
%\begin{align}\label{eq:overlap}
%    &\langle \Psi^t_{z_1}, \Psi^t_{z_2}\rangle=\\
%    &=\left(\frac{X\sqrt{\sinh(\eta_1)\sinh(\eta_2)}}{2\sinh(X/2)\sqrt{\eta_1\eta_2}}e^{\frac{X^2/2-\eta_1^2-\eta_2^2}{2t}}\right)^{3M^3},\nonumber
%\end{align}
%where $X= -i(\xi_1-\xi_2)+\eta_1+\eta_2$. Consequently, 
 if $z_1\neq z_2$, for $t\to 0$ we enjoy exponential decay of the overlap, i.e. $\langle \Psi^t_{z_1}, \Psi^t_{z_2}\rangle= \mathcal{O}(e^{-1/t})$.

%Further, inside of an expectation value $\langle .\rangle := \langle \Psi^t_{(c,p)}| \,.\, |\Psi^t_{(c,p)}\rangle$ we are allowed to replace in monomial operators the Ashektar-Lewandowski volume,
%\begin{align}
%&\hat V^{AL} _v = \frac{(\beta\hbar\kappa)^{3/2}}{2^{7/2}\sqrt{3}}\sqrt{|\hat Q_v|},\\
%&\hat Q_v = (\frac{i\,\hbar\kappa\beta}{2})^{3} \sum_{ijk}\epsilon(i,j,k)\epsilon_{IJK} \hat P^I(e_i)\hat P^J(e_j)\hat P^K(e_k),\nonumber
%\end{align}
%(where $\epsilon(i,j,k):=sgn(\det(\dot{e}_i,\dot{e}_j,\dot{e}_k))$) with the polynomial $k$-th Giesel-Thiemann volume,
%\begin{align}\label{giesel-thiemann-volume}
%&\hat{V}^{GT}_{k,v}=\frac{(\beta\hbar\kappa)^{3/2}}{2^{7/2}\sqrt{3}}\sqrt{\langle \hat Q_v\rangle}\;\times\\
%&\times\left[\id_{\Hil}+\sum_{n=1}^{2k+1}\frac{(-1)^n}{n!}\left(n-\frac{5}{4}\right)!\left(\frac{\hat Q_v^2}{\langle \hat Q_v\rangle^2}-\id_{\Hil}\right)^n\right],\nonumber
%\end{align}
%making only a mistake of order $\mathcal{O}(t^{k+1})$ \cite{xxxx}. {\bf [All of this volume stuff is probably not needed]}\\

Here, we are also able to pose quantitative statements about the variance of the Thiemann-regularization of the Euclidean part of the scalar constraint with the lapse function fixed at unity
\begin{align}\label{CE_operator}
    \hat{C}_E:=C_E (\{\hat h(e)\}_{e\in\gamma},\{\hat P(e)\}_{e\in\gamma})
\end{align}
which is a complicated expression in terms of fluxes and holonomy operators, see \cite{LR:20} for a common regularisation.
%\begin{align}\label{CE_operator}
%    &\hat{C}_E[1]:=p_{\rm Eucl}\sum_{v\in\gamma} \sum_{ijk\in L}\left(\hat h(\square^{\epsilon}_{v,ij})-\hat h^\dagger(\square^\epsilon_{v,ij})\right)\times\nonumber\\
%    &\;\;\;\times \epsilon(i,j,k)  \hat h(e_k) \left[\hat h^\dagger(e_k), \hat{V}^{GT}_{2,v}+\hat{V}^{GT}_{2,v+e_k}\right]+cc.
%\end{align}
%We denote $L=\{\pm1,\pm2,\pm3\}$ and $p_{Eucl}=1/(i\kappa^2\beta \hbar)$ while the volume $\hat V^{GT}$ is a polynomial function of right-invariant vector fields. 
Primarily, this is possible due to a very recent and highly non-trivial result concerning the expectation values of monomials in the operators (\ref{eq:holonomy}) and (\ref{eq:flux}) on a single edge, $\langle\hat{h}^{(j)}_{ab}R^{K_1}...R^{K_N}\rangle_{\psi^t_{G_I}}$, where next-to-leading order is included \cite{LZ:19}. We postpone exact, compact
formulas to the last section of Supplementary Material. Building on this, a feasibly implementable algorithm which allows to compute expectation values in $\Psi^t_{z}$ including the next-to-leading order corrections for arbitrary (polynomial) operators has very recently been introduced and tested \cite{LR:20}. A detailed description of the algorithm can be found in Sec. 4 of \cite{LR:20}. 

With this algorithm it is possible to compute the expectation values of nested operators. In particular, the expectation value of $\hat C_E$ has been reported in \cite{LR:20}\footnote{
An implementation of the algorithm as a free-to-use Mathematica code is available at \cite{github}.
}. Here, we use this algorithm to compute such a complex object as $\Delta_{\Psi^t_{z}}\hat C_E$. % --- see Eq. (\ref{Variance:CE}) below --- obtaining more fundamental information about the dynamics of $\Psi^t_{z}$.
%
 %However, as our algorithm \cite{LR:20} in principle allows to compute the variance of the Hamiltonian constraint, Eq. (\ref{qsl}) becomes practically meaningful. In particular, 
 Our implementation \cite{github} applied to $\hat C_E$, defined in (\ref{CE_operator}) with the lapse function fixed at unity, gives the result
\begin{align}\label{Variance:CE}
    &\left(\Delta_{\Psi^t_{z(0)}} \hat{C}_{E}\right)^2 = \frac{3}{2^7}
    \frac{M^3  \beta\hbar}{\eta \kappa}\sin(\xi)^2\times\\
    &\times \left(17+256\eta^2+(256\eta^2-17)\cos(2\xi)\right)+\mathcal{O}(t). \nonumber
\end{align}
We focus on $\hat C_E$ for the sake of simplicity, as our goal is to convey a general message about usefulness of QSL in LQG. We also admit that further optimization of the algorithm's implementation is necessary to compute the variance of the full scalar constraint. 

\textit{QSL for Euclidean LQG}---Now, all the tools are at hand to apply the QSL to Quantum Gravity. As the cosmological coherent states are the prime proposal for the semi-classical limit of LQG, we will test their validity with respect to the Euclidian part of the scalar constraint. To this end we fix the family of states present in Eq. (\ref{eq:Wapprox1}) to be $\Phi_s=\Psi^t_{z(s)}$, and let the generator be $\hat H=\hat C_E$. Note that while numerous applications of QSL are known in the literature \cite{PhysRevLett.110.050402,PhysRevLett.110.050403,PhysRevLett.120.070401,PhysRevLett.120.070402,PhysRevLett.120.060409,Campaioli2019tightrobust,PhysRevA.103.022210,PhysRevResearch.2.023125,PhysRevLett.126.180603},
%\footnote{Most famously, for the quantum Harmonic oscillator and Gaussian states \Red{....}} 
this tool has never before been introduced to Quantum Gravity.

In fact, for the family of coherent states (\ref{eq:coh_state_def}), whose overlap is known to decrease exponentially fast with $t\to 0$ if $z\neq z'$, the necessary condition of the QSL can easily be violated. Given a trajectory $z(\tau)$, such that $\lim_{t\to 0} [z(0)- z(\tau)]\neq 0$ for all $\tau>0$, we get $\lim_{t\to 0} A(0,\tau) = \pi/2$. Moreover, since $\lim_{t\to 0} \Delta_{\Psi^t_{z(0)}} \hat C_E <\infty$, then for $\tau<<1$ the negative term on the right hand side of (\ref{qslrev}) is negligible. Consequently, $W(\tau,\tau)\approx 0$, indicating that the state almost immediately leaves the desired trajectory, becoming approximately orthogonal to the state $\Phi_\tau$. Hence, the family of such coherent states is not stable.
%Proof: Let $C_0:= \lim_{t\to 0} \Delta_{\Psi^t_{z(0)}} $, which by assumption is finite. Then choose $\tau>0$ s.t. $\tau C_o < \delta$ for some $\delta>0$. By assumption there exist $\epsilon>0$ s.t. $\lim_{t\to 0} z(\tau) - z(0) > \epsilon$, hence $\lim_{t\to0}W(0,\tau) = 0$. Thus, for any $\delta'>0$ we find $t_o<<1$ such that W_{t_o}(0,\tau)<<\delta'$. It follows $\delta> |\arcos(W_{t_0}(\tau.\tau))-\arcos(\delta')|$. Hence it must be W_{t_0}(\tau.\tau)\approx 0$.

While above we have shown that there is a serious risk of violation of (\ref{qslrev}), we now %use (\ref{Variance:CE}) to 
prove that, at least for very short times, this risk is mitigated by {\it natural} choice for $z(\tau)$ which parametrizes the semi-classical family.
%Recall: here M^3 is the total number of lattice points, while in mathematica file it was only M
To this end, we write $\eta(\tau)= 2 t p(\tau)/(M^2\hbar \kappa\beta)$ and $\xi(\tau)=c(\tau)/M$, where $\left(p(\tau),c(\tau)\right)$ is a trajectory on the 1-dimensional phase space $\{p,c\}=\kappa\beta/6$, rendered by the generator $C_E(p,c)=(6M^2/\kappa)\sqrt{p}\sin(c/M)^2$ being a leading order of the average of $\hat C_E$ evaluated on flat isotropic cosmology.

 Looking at very short times %\footnote{Note, that the limits of $\tau\to0$ and $t\to0$ do not commute, thus making both distinct investigations necessary.}
$\tau<<t$  we get (see the second section of Supplementary Material)
\begin{align}\label{eq:qsl2}
   &\left(\partial_\tau A(0,\tau)|_{\tau=0}\right)^2 = \frac{3}{2^7}\frac{M^3\beta}{\eta \kappa\hbar}
    \sin(\xi)^2\times\\
    &\;\;\times  (16+256\eta^2+(256\eta^2-16)\cos(2\xi))+\mathcal{O}(t)\nonumber\\
   \equiv&  \left(\Delta_{\Psi^t_{z(0)}} \hat{C}_{E}\right)^2/\hbar^2-\frac{3}{2^6}
    \frac{M^3  \beta}{\eta \kappa\hbar}\sin(\xi)^4+\mathcal{O}(t).\nonumber
\end{align}
%where the last line stems from a comparison with (\ref{Variance:CE}). 
Since by definition $A(0,0)=0$ while %it can be checked that independent on the choice of $M$ and $\beta$ we find $\forall \eta>0, \xi\in [0,\pi)$ that 
Eq. (\ref{eq:qsl2}) implies $ \Delta_{\Psi^t_{z(0)}} \hat{C}_{E}/\hbar\,\geq\,\partial_\tau A(0,\tau)|_{\tau=0} $, we fascinatingly observe that the right hand side of (\ref{qslrev}) is negative for very short times. Hence, evolution due to $\hat C_E$ obeys  the necessary condition for sharp peakedness, i.e. $W(\tau,\tau)\approx 1$ is not automatically ruled out. In other words, the time to transfer from one coherent state to the next one is sufficient, at least for $\tau<<t$. However, albeit the QSL necessity check is passed, we point out that stability of the GCS family is not automatically proven.

\textit{Further applications}---
While the main message of this letter is the consistency check for semi-classical families, the potential for applying QSL does not end here. E.g., the QSL in Eq. (\ref{qsl}) can be used to make further conclusions for sharply peaked states with $t << \tau$, which are peaked on huge fluxes. By the latter we mean that $p(0)$ and $c(0)$ scale with $t$ in such a way that $\xi(0), \eta(0)\neq 0$ are both independent of $t$. %If not, then \lim_t \eta(o) \to 0. In which case the equations are not well defined anymore due to 1/\eta \to \infty.
 The regime $t<<1$ then corresponds to the semi-classical limit with huge fluxes and small relative variances. Now, there exists $\tau$ making the left hand side of (\ref{qsl}) arbitrary small. Hence, as already explained, the stability directly depends on $W(0,\tau)$, and due to orthogonality of GCS there is a risk that $W(0,\tau)\approx 0$. 

However, for the toy model driven by $\hat C_E$ we show at the end of the second section of Supplementary Material that $z(\tau)=\tilde z(\sqrt{t}\tau)$, with $\tilde z$ otherwise independent of $t$. This implies that effectively the flow parameter rescales with $\sqrt{t}$ and the evolution freezes in the limit $t\to0$. Thus, for fixed $\tau$ we get both $W(0,\tau) \to 1$ and $W(\tau,\tau)\to 1$. For short times we do have a family with infinitely huge fluxes, which in the limit $t\to 0$ is stable. 
Further details on when such a rescaling occurs for other operators, as well as its implications for finite-time stability of generalized coherent states can be found in \cite{Kaminski:20,Kaminski:TBA}.

\textit{Conclusions.}---In recent years, the conceptual side of LQG has developed a lot. However, most if not virtually all practical questions in the field require to perform computations of drastically huge complexity. 
As a part of efforts directed towards this objective of making complex objects in LQG computable, we have recently established an algorithm \cite{LR:20} offering a severe simplification of the computations with the coherent states in LQG. In this letter we employ these complex techniques to explore new physics within LQG. In particular, our methodology, for the first time, allows to probe the evolution of the coherent states. To this end, we brought quantum speed limit, usually discussed within standard quantum theory, and after slight adjustments applied it to the evolution of the coherent states. However, the inequality offered by the quantum speed limit is informative only if one is able to compute or measure the variance of the generator of the evolution. Since we show that with the techniques developed, it is now possible to access such a complex quantity, a thorough discussion of the physics underlying the evolution of the coherent states is possible. The physical problem at hand, which is the main theme of this letter is the following. We know that a coherent state will evolve to a state which is not exactly coherent. On the other hand, classical evolution on the phase space, by construction, would mean that in the quantum description we stay in the family of coherent states, just their parameters evolve appropriately. Then, what is the deviation between such quantum and classical evolution? With this letter we provide a first attempt towards answering such a question, which from a computational perspective is extremely difficult. Interestingly our finding is that the deviation between both states, at least for short times, is infinitesimal.  This conclusion, \textit{a priori} is neither obvious nor even expected, due to the fact that the overlap between the coherent states decays exponentially. With more computational effort it will be interesting to extend these considerations to the various quantum regularizations of the Lorentzian part.

We have seen that the QSL can serve as a powerful tool to rule out candidates for semi-classical families and indicate those for which stability is still possible.
 The findings stemming from our methodology also raise the hope for full LQG to align with previous investigations in LQC, where the only genuinely quantum investigations happened so far \cite{Bojo:05}. 
In particular, by  minimally coupling isotropic gravity to a free scalar field, which could serve as a clock, the authors of \cite{Ash:06,APS:06_b,APS:06_c} deparametrized the system to obtain an evolution operator on a 1-particle Hilbert space. In this way, they were maximally close to the setting described above and, after rigorous analytical and numerical work, it transpired that standard Gaussian states remain stable under such evolution \cite{Ash:06,APS:06_b,APS:06_c,ADLP:18}. In the language of the current manuscript, these states would obviously pass the QSL test. These results nurtured the conjecture that the same might be true for more elaborate models. However, due to a tremendous complexity, this conjecture has not only never been tested, but also feasible tools to probe it were basically missing. Our work opens up the latter possibility. It also allows one to follow up on earlier investigations such as \cite{Kuchar89}, enabling tests of minisuperspace models. A first consistency check of the assumed stability in these models can be an attractive starting point for future work.

In fact, pinpointing a suitable candidate for semi-classical models will create a crucial fundamental for the years to come: it is predicted that LQG will soon start to enter the realm of quantum simulations \cite{Feynman, Lloyd}. Several platforms, for example those utilizing linear optics \cite{optical}, adiabatic (quantum) computation \cite{adiabatic}, Nuclear Magnetic Resonance techniques \cite{NMR} or  superconducting qubits \cite{superconducting} are considered. Hence, before the era of quantum simulations in LQG, preparatory work - like in this letter -aiming at making complex objects in LQG computable (by both classical and quantum machines) shall be in focus.

%Last but not least, since this manuscript deals with the dynamical behaviour of the cosmological sector for Quantum Gravity, appropriate quantization of the constraints shall precede any discussion. In this letter we recalled two standard prescriptions and added a third version, namely, the weak implementation. In this approach one only demands that the measurements of the constraint amount to "zero plus infinitesimal" quantum fluctuations. From a philosophical perspective, this implementation of the constraints is favourable for investigations of semi-classical systems. 

%Comment: continuum limit cannot be taken for the obtained results (such as stable states, due to giesel-thiemann volume.

\acknowledgements
The authors want to thank Wojciech Kami\'nski for clarifying discussion on the quantum speed limit in the regime of huge fluxes% and Laura Herold for pointing out the possibility to simplify the measure in (\ref{eq:ToyModel_Mu})
. 
{\L}.R. would like to acknowledge support by the Foundation for Polish Science
(IRAP project, ICTQT, Contract No. 2018/MAB/5, cofinanced by the EU within the Smart Growth Operational Programme).

\bibliography{prl-bib}{}
\bibliographystyle{ieeetr}

\clearpage
\newpage

\begin{widetext}
\section{Supplemental Material for \textit{Quantum speed limit and stability 
of coherent states in quantum gravity}}
\end{widetext}

\section{A generalized derivation of quantum speed limit}
Even though derivation of quantum speed limit can easily be found in the literature (consult, for example, a  recent pedagogical review \cite{Campaioli}), we shall repeat it here because in the main text we use a non-standard variant of the QSL. Perhaps, the inequality to be presented, even though conceptually as simple as the typical QSL, was never in the focus due to a more general assumption we shall make concerning one of the quantum states in question. 

%and apply it exemplarily to the quantum harmonic oscillator, whose dynamical stability is well-known.\\

One starts with the Heisenberg uncertainty relation for two operators $\hat A$ and $\hat H$
\begin{align}
    \Delta_{s} \hat A \Delta_{s} \hat H \geq \frac{1}{2} |\langle \Phi(s) , [\hat A,\hat H] \Phi(s)\rangle |,
\end{align}
one of which being the Hamiltonian. We simplify the notation, so that $\Delta_{s} $ denotes $\Delta_{\Phi(s)} $, where $\Phi(s)=e^{-i\hat H s/\hbar} \Phi_0$ is the evolved initial state $\Phi(0)\equiv \Phi_0$. Clearly, $\Delta_s \hat H = \Delta_0 \hat H$, so that it does not depend on $s$.

Since $\partial_s\Phi(s)=-\frac{i}{\hbar}\hat H \Phi(s)$ it follows that

\begin{align}
     \Delta_{s} \hat A \Delta_{0} \hat H \geq \frac{\hbar}{2} | \partial_s \langle \Phi(s) , \hat A  \Phi(s)\rangle |.
\end{align}
From now on, everything depends on the choice of $\hat A$. In the standard treatment $\hat A$ is taken to be the projector on the \textit{initial} state $\Phi_0$. In such a way it is possible to track the evolution from the initial to the final state. In our treatment we generalize this approach and let $\hat A$ be the projector on an \textit{arbitrary} state $\Upsilon$. Still, we have that $\hat A^2=\hat A$. Therefore, we can easily rewrite the previous inequality into
\begin{align*}
    \Delta_0 \hat H \geq \frac{\hbar}{2}\frac{|\partial_s \langle \Phi(s) , \hat A  \Phi(s)\rangle|}{\sqrt{\langle \Phi(s) , \hat A  \Phi(s)\rangle\left(1-\langle \Phi(s) , \hat A  \Phi(s)\rangle\right)}}.
\end{align*}
Since
\begin{equation*}
    \frac{1}{2}\int_0^\tau ds \frac{\partial_s f(s)}{\sqrt{f(s)(1-f(s))}}=\arccos(\sqrt{f(0)})- \arccos(\sqrt{f(\tau)}),
\end{equation*}
we can integrate both sides of the above inequality with respect to $s$ from $0$ to $\tau$, enter with the integral inside the absolute value on the right hand side, and obtain
\begin{widetext}
\begin{align}\label{qslpart}
    \tau\cdot \Delta_0 \hat H/ \hbar \geq \left| \arccos\left(\sqrt{\langle \Phi_0 , \hat A  \Phi_0\rangle}\right) - \arccos\left(\sqrt{\langle \Phi(\tau) , \hat A  \Phi(\tau)\rangle}\right) \right|.
\end{align}
\end{widetext}
This is a more general variant of QSL. As already mentioned, the standard QSL follows if $\Upsilon=\Phi_0$, so that $\langle \Phi_0 , \hat A  \Phi_0\rangle=1$, and consequently the first term inside the absolute value vanishes. Furthermore, in the special case in which the final state $\Phi(\tau)$ is orthogonal to the initial state $\Phi_0$, the right hand side of (\ref{qslpart}) equals $\pi/2$. In that case, the transition time between the two orthogonal states is at least \cite{Mandelstam1991}
\begin{align}
    \tau \geq \frac{\hbar \pi }{2\Delta_0 \hat H}.
\end{align}

Finally, let us observe that our QSL in the main text follows from (\ref{qslpart}) if we set  $\Upsilon=\Phi_\tau$. While $\Phi_0$ remains to play a role of the initial  state, $\Upsilon$ is neither an initial nor final state of the evolution, i.e. neither $\Phi_0$ nor $\Phi(\tau)$. Instead, $\Upsilon$ represents some \textit{target} state $\Phi_\tau$, i.e. the state we wish to obtain during the evolution. 

\section{Derivation of Eq. (14) in the main text and further details}
We first recall that the classical Euclidean part of the scalar constraint is
\begin{align}\label{supeq:classical_constraint}
    C_E(p,c)= \frac{6M^2}{\kappa}\sqrt{p}\sin(c /M)^2,
\end{align}
while the Poisson bracket on the reduced phase space reads
\begin{align}
    \{p,c\}=\kappa\beta /6.
\end{align}
Using the Hamilton's equations of motion we can thus expand the solution for short times:
\begin{align}
    p(\tau)&\approx p(0) + \tau \; \{C_E,p\} + \frac{\tau^2}{2} \{C_E,\{C_E,p\}\}=\\
    &=p-M\beta\sqrt{p}\sin(2c/M)\tau+\beta^2\frac{1-\cos(2c/M)}{4/M}\tau^2,\nonumber
\end{align}
with $p\equiv p(0)$, and similarly for $c$. Translating this solutions to the variables of the GCS
\begin{align}\label{supeq:chocie_z}
    \eta=\frac{2t}{M^2\hbar\kappa\beta}p,\;\;\; \xi =c/M,
\end{align}
leads to its respective expansion for short times
\begin{align}\label{eta_expansion}
    &\eta(\tau)\approx\eta -\sqrt{\frac{2\beta  \eta  t}{\kappa  \hbar }} \tau  \sin (2 \xi ) +\beta  t \tau ^2\frac{1- \cos (2 \xi )}{2 \kappa  \hbar },
\end{align}
\begin{align}
    &\xi(\tau)\approx \xi +\frac{t  \sin ^2(\xi )}{ \sqrt{2\eta t\kappa\hbar/\beta}}\tau +\frac{\beta  t\sin ^3(\xi ) \cos (\xi )}{\eta  \kappa  \hbar } \tau ^2. \label{xi:expansion}
\end{align}
Let us note in passing that in this way we also obtain an alternative form of the Poisson bracket
\begin{align}\label{supeq:alt_PB}
    \{\eta,\xi\}= t/(3\hbar M^3).
\end{align}
With (\ref{eta_expansion}) and (\ref{xi:expansion}) we obtain $z(\tau)=\xi(\tau)-i \eta(\tau)$. We use this variable in order to compute
\begin{widetext}
\begin{align}
    &|\langle \Psi^t_{z(0)}, \Psi^t_{z(\tau)}\rangle|=\cos\left(A(0,\tau)\right)=\\
    &1-\frac{3 \beta  \tau^2 \text{csch}^2(\eta ) \sin ^2(\xi ) \left(\left(16 \eta ^2-1\right) \cos (2 \xi )+16 \eta ^2+1\right) \left(-2 \eta ^2+2 \eta ^2 t-\cosh (2 \eta ) \left(t-2 \eta ^2\right)+t\right)}{64 \eta ^3 \hbar \kappa /M^3}+\mathcal{O}(\tau^3).\nonumber
\end{align}
\end{widetext}
In the next step we take $\arccos(\cdot)$ in order to arrive at an approximate form of $A(0,\tau)$. Lastly, we take the limit  $t\to 0$ to get the leading order expansion:
\begin{align}
    &A(0,\tau)= A(0,0) + \tau  (\partial_\tau A(0,\tau))|_{\tau=0} +\mathcal{O}(\tau^2),\\
    &A(0,0)=0,\\
    &(\partial_\tau A(0,\tau)|_{\tau=0})^2 = \frac{3}{2^7}\frac{M^3\beta}{\eta \kappa\hbar}
    \sin(\xi)^2 \\
    &\quad\quad\quad\left(16+256\eta^2+(-16+256\eta^2)\cos(2\xi)\right)+\mathcal{O}(t).\nonumber
\end{align}
Note that $W(0,\tau)$ simply follows from the definition of the overlap.

We can also use Eqs. (\ref{eta_expansion}) and (\ref{xi:expansion}) to re-obtain the equations of motion written in terms of $\eta$ and $\xi$:
\begin{align}
    &\frac{\partial\eta}{\partial \tau}= -\frac{\sqrt{2\beta\eta}}{\sqrt{\hbar\kappa/t}}\sin(2\xi)\\
\end{align}
\begin{align}
    &\frac{\partial\xi}{\partial \tau}=\frac{\sqrt{\beta}}{\sqrt{2\eta\hbar\kappa/t}}\sin(\xi)^2
\end{align}
While making a substitution $\tilde{\tau}=\tau \sqrt{t}$, we can see that the equations become independent of $t$. Given the initial conditions $\eta(0),\xi(0)$ we therefore get a unique solution $\xi(\tau \sqrt{t}), \eta(\tau \sqrt{t})$. Consequently, we find that in the limit $t\to 0$ the evolution freezes.

\section{Expectation values}
Here we collect the expectation values mentioned in the main text, which are a vital ingredient of the algorithm presented in Sec. 4 in \cite{LR:20}.
Working in the spherical basis $\uptau^{K}$ with $K_1,...,K_N \in \{\pm 1, 0\}$, we have  \cite{LZ:19}:
\begin{widetext}
\bal
\langle&\hat{h}^{(j)}_{ab}R^{K_1}...R^{K_N}\rangle_{\psi^t_{G_I}} = \langle1\rangle \left(\frac{i\eta}{t}\right)^N  D^{(1)}_{-K_1-S_1}(n_I)...D^{(1)}_{-K_N-S_N}(n_I)\sum_{c=-j}^j \; \Bigg(\delta^{S_1}_0...\delta^{S_N}_0\delta_{aa'}+\label{hRRR}\\
&+\frac{t}{2\eta}\bigg[\delta_{aa'}\delta^{S_1}_0...\delta^{S_N}_0
N\left(\frac{N+1}{2\eta}-{\rm coth}(\eta)\right)+i\sum_{A=1}^N \delta^{S_1}_0...\cancel \delta^{S_A}_0 ...\delta^{S_N}_0 \left(1-s_A {\rm tanh}\left(\frac{\eta}{2}\right)\right)D^{(j)}_{-s_A-L}(n^\dagger_I)[\uptau^{L}]^{(j)}_{aa'}\nonumber\\
&\left.-\frac{\delta_{aa'}}{\sinh(\eta)}\sum_{A<B=1}^N\delta^{S_1}_0...\cancel \delta^{S_A}_0...\cancel \delta^{S_B}_0... \delta^{S_N}_0(\delta^{S_A}_{+1}\delta^{S_B}_{-1}+\delta^{S_A}_{-1}\delta^{S_B}_{+1})e^{S_A\eta}\bigg]
\right) \; D^{(j)}_{aa'}(n_I) e^{-i\xi c} \gamma^j_c D^{(j)}_{cb}(n^\dagger_I)
\;+\mathcal{O}(t^2),\nonumber
\eal
\end{widetext}
with $\gamma^j_c=1-t \tilde\gamma^j_c$, where \cite{LZ:19}
\bal\label{GammaResult}
\tilde\gamma^j_c=\frac{1}{4}\left[(j^2+j-c^2)\frac{{\rm tanh}(\eta/2)}{\eta/2}+c^2\right].
\eal
As in the main text, by $D^{(j)}_{mn}$ we denote Wigner matrices in the $j$th representation. Moreover, the overlap reads
\bal\label{normalisation}
\langle 1\rangle:&=\langle \hat{h}^{(0)}_{00}\rangle_{\psi^t_{G_I}}=
\sqrt{\frac{\pi}{t^3}}\frac{2\eta \,e^{\eta^2/t}}{\sinh(\eta)}e^{t/4}.
\eal

\end{document}